\documentstyle[aps,prb,multicol]{revtex}
\begin{document}
\draft
%
\hoffset= -2.5mm

\title
{Comment on ``Anomalous crossover between thermal and shot noise
in macroscopic diffusive conductors''}

\author{Mukunda P. Das and Frederick Green\\}
\address{
Department of Theoretical Physics,
Research School of Physical Sciences and Engineering,
The Australian National University,
Canberra ACT 0200, Australia}

\maketitle

\begin{abstract}
Shot noise is not normally evident in bulk solid-state conductors,
since it is strongly attenuated by inelastic collisions.
The ``anomalous'' emergence of macroscopic shot noise is discussed
in G. Gomila and L. Reggiani, Phys. Rev. B {\bf 62}, 8068 (2000).
We remark on the consistency of this linear diffusive
model at the large voltages and currents needed to see the effect.


\end{abstract}

\begin{multicols}{2}

Gomila and Reggiani
\cite{gr}
(GR) have revived the possibility of observing
shot noise in a macroscopic conductor.
The electronic properties of the medium may range
from strongly degenerate to classical.
If the device length is much greater than
the mean free path
for carrier scattering, we are in the bulk regime.

At high enough currents GR predicts a
novel and different type of smooth crossover
from thermal noise to shot noise. This
crossover does not develop in the way
established for elastic scattering at mesoscopic scales
\cite{blbu}.
Under normal conditions,
macroscopic solid-state shot noise is completely
extinguished by inelastic collisions.
Dissipative inelastic processes
dominate transport at bulk scales,
as also at large applied potentials.

Central to GR is the idea that
elementary carrier-number fluctuations are
directly responsible for shot noise.
This proposition is a step forward in understanding,
and one that departs from the accepted wisdom
\cite{kogan}
for solid-state noise.
That theoretical wisdom precludes any distinction
between carrier-number fluctuations, as such,
and fluctuations of the carriers' free energy
(associated with thermal noise).

Shot noise as the unique signature of
number fluctuations is not new.
It has been formulated within a conventional and rigorous
kinetic-equation framework for high-field noise
\cite{gd1,gd2,gd3,gd4,gd5,gd6}.
Quite unrelated to that perspective is
Naveh's drift-diffusive approach
\cite{naveh}
to the rise in macroscopic shot noise.
His work well predates GR.

Despite the advocacy of a number-fluctuation origin of
shot noise, other notions put forth in Reference {\onlinecite{gr}}
call for some caution. Below, we
address (a) the validity of a linear model;
(b) the zero-temperature behavior of shot noise; and
(c) the relevance of Coulomb correlations.
The internal consistency
of each of these, within GR, is problematic.
Additional issues, such as the makeup of the diffusion
equation and Langevin stochastics for degenerate systems,
are covered in Ref. \onlinecite{gd6}.

(a) {\bf Applicability of the formalism}.
The linear drift-diffusive analysis of GR is intrinsically
tied to weak external fields and low currents.
It cannot be applied at high fields,
just where bulk shot noise is said to emerge.

An estimate of the field required for
the onset of anomalous noise
is made in Ref. \onlinecite{gr}
for a nondegenerately doped semiconductor
at a carrier density of $10^{14}{~}{\rm cm}^{-3}$.
The value comes out above $2 {~}{\rm kVcm}^{-1}$.
On the other hand, the upper limit of validity
for hydrodynamic diffusion is easy to derive.
That field determines the point at which the
energy transferred in inelastically mediated {\it drift}
exceeds the scale typical of dynamical {\it diffusion}.
At this point the linear Einstein relation,
on which the diffusion equation depends, breaks down.

The upper limiting field $E_{d}$
for drift-diffusion equivalence is given approximately by

\begin{equation}
eE_{d} \lambda_{in} \lesssim \hbar v_{th}/\lambda
\label{eq1}
\end{equation}

\noindent
where $\lambda_{in}$ and $\lambda$
are the inelastic-only and net mean free paths respectively,
and $v_{th}$ is the thermal velocity.
This simply compares the energy gained in drifting
across $\lambda_{in}$ with the energy associated
with diffusion in the time $\lambda/v_{th}$.
At temperature $T = 300$ K in, say, GaAs we have
$v_{th} \sim 10^7{~}{\rm cms}^{-1}$ and
$\lambda \approx \lambda_{in} \sim 50$ nm.
This yields

\begin{equation}
E_{d} = 25{~}{\rm Vcm}^{-1}
\ll 2\times10^3{~}{\rm Vcm}^{-1}.
\label{eq2}
\end{equation}

\noindent
Hence the regime for anomalous shot noise, as predicted
by GR, lies beyond the reasonable application of
linear diffusive response. As a way of describing
the far-from-equilibrium domain (the declared goal of GR
\cite{gr})
linear theory is inadequate.

There seems to be an implicit assumption here that drift-diffusion,
plus knowledge of the low-field conductance, are enough to
determine nonequilibrium noise.
However, the specifics of one-body transport, linear or not
(cf. comment 13 of Ref. \onlinecite{gr}),
shed no light on the essentially {\it two-body} structure
of fluctuations far from equilibrium.
That has been strikingly shown by Stanton and Wilkins
\cite{sw}.

(b) {\bf Temperature scaling}.
Equations (10) and (16) of Ref. \onlinecite{gr}
are said to determine how large-scale shot noise emerges.
This is the anomalous crossover in the bulk.
Both equations describe a noise spectral density
$S_I$ that subsumes thermal and anomalous effects.
Overall, $S_I$ would appear to scale
with ambient temperature through
its leading factor, proportional to the
Johnson-Nyquist spectrum $S^{ther}_I = 4k_BTG$
where $G$ is the conductance.

Shot noise, unlike Johnson-Nyquist noise,
remains well-defined and finite as $T \to 0$.
To recover this definitive characteristic,
Eqs. (10) and (16) of GR must generate an anomalous term
bearing a compensating denominator that scales with $T$
to undo the leading factor $S^{ther}_I$.
Otherwise, the outcome is just $S_I\propto T$.
This shot noise would vanish in the zero-temperature limit,
exactly like any other thermally induced fluctuations.

In GR the ratio $S_I/S^{ther}_I$ is a function
of three lengths operative in the problem
\cite{gr}.
For metallic electrons, straightforward inspection of Eq. (10)
shows that the ratio carries no counter-term for
the leading factor $S^{ther}_I$ as $T \to 0$. That is,

\begin{equation}
\lim_{T \to 0} {\{ S_I - S^{ther}_I \}} \sim S^{ther}_I \to 0.
\label{eq3}
\end{equation}

\noindent
Thus the ``anomalous'' component of GR is a special
case of excess {\it thermal} noise.

All accepted descriptions of mesoscopic noise
result in the standard crossover formula
\cite{blbu,kogan}
(cited as Eq. (2) in Ref. \onlinecite{gr}).
Its current-fluctuation behavior
evolves from low-field thermal noise $S^{ther}_I$
to a nonequilibrium shot-noise asymptote that
(i) scales with the classical value $2eI$ and (ii) displays
absolutely {\it no} dependence on $T$.

Both (i) {\it and} (ii) define the crossover.
Both follow naturally if shot noise is produced by
randomly timed carriers, discretely perturbing
the population as they enter and leave
through the conductor-lead interfaces
\cite{ldr93}.
For GR, one would have expected an asymptote
going as $2eI$ independently of $T$.
This is evidently not the case.
Whatever is meant by the label ``anomalous''
\cite{gr}
for metallic electrons, the behavior of $S_I$
is unavoidably linear in $T$.

Thermal sensitivity is not a feature of true shot noise
\cite{dang}.
GR's only kinship with the standard crossover, as
described by every drift-diffusive theory,
is its high-current linear trend.

(c) {\bf Bulk Coulomb effects}.
Gomila and Reggiani study conductors that are
large, homogeneous, and in which Thomas-Fermi screening
\cite{pinoz}
is strong even at moderate degeneracy. For example,
in silver the screening length is $\lambda_{\rm TF} = 0.1$ nm;
in Si-doped GaAs at $10^{18}{~}{\rm cm}^{-3}$ we have
$\lambda_{\rm TF} = 10$ nm. Already well below any
mean free path, these values are minute compared with
the sample size.
It would be surprising to see marked Coulomb effects
in the bulk, especially at frequencies negligible
with respect to the plasmon resonance. 

Gomila and Reggiani argue for a pronounced departure
from the accepted crossover, arising from
the interplay of Coulomb-induced fluctuations
and carrier transport (both diffusive and ballistic).
That inevitably implies the presence of internal-field
correlations stretching over macroscopic distances.
Is this scenario a likely one?

The reason why it is not lies in the principle
of quasineutrality, or perfect screening
\cite{pinoz}.
A {\it uniform} electron gas cannot support
deviations from local neutrality outside the range
of  $\lambda_{\rm TF}$.
Quasineutrality in a uniform system persists
away from the equilibrium limit
\cite{gd5}.

Microscopically, current fluctuations are made up of
polarized electron-hole (e-h) excitations. The conservation laws
require an intimate kinematic linkage
between e-h components of a polarized pair
\cite{pinoz}.
Because drift-diffusive analysis
denies the excited electron-hole partnership,
it totally misses the physics of e-h pair propagation
and screening
\cite{gd2,gd3,gd4}.
Thus GR also contravenes microscopic quasineutrality.

Diffusive models fail to respect electron-hole symmetry.
Fulfillment of perfect screening is out of the question
here.
One sees this unphysical asymmetry even at the
coarser-grained level of one-body transport;
note, for example, Fenton's critique
\cite{fenton}.

A full kinetic-equation treatment of the ``anomalous'' problem,
conforming explicitly to the physics of the electron gas
\cite{gd4,gd5},
correctly accounts for
the {\it temperature-independent} nature of
shot noise in metallic wires, be they long or short
\cite{gd1,gd6}.
This behavior -- not seen in GR -- is
typical of carrier-number fluctuations.
Moreover, we predict {\it strong suppression}
of high-field shot noise
in wires whose diameter is less than the mean free path.
That effect awaits experimental investigation.

In sum: we have shown why the
conclusions of Gomila and Reggiani
\cite{gr},
on the emergence of bulk high-field shot noise,
should be regarded cautiously.
On the other hand, a simple and exact
nonequilibrium kinetic calculation
\cite{gd1,gd6}
suffices to quantify bulk shot noise
even for {\it noninteracting} carriers
(say at density $10^{14}{~}{\rm cm}^{-3}$,
as in GR's proposed test of the
Coulomb-induced crossover anomaly).

\end{multicols}


\begin{references}

\bibitem{gr}
G. Gomila and L. Reggiani, Phys. Rev. B {\bf 62}, 8068 (2000).

\bibitem{blbu}
Ya. M. Blanter and M. B\"uttiker, Phys. Rept. {\bf 336}, 1 (2000).

\bibitem{kogan}
Sh. M. Kogan, {\it Electronic Noise and Fluctuations in Solids}
(Cambridge University Press, Cambridge, 1996), Ch. 5.

\bibitem{gd1}
F. Green and M. P. Das, preprint cond-mat/9809339.

\bibitem{gd2}
F. Green and M. P. Das, in {\it Proceedings of the Second
International Conference on Unsolved Problems of Noise
and Fluctuations (UPoN'99)}, ed. D. Abbott and L. B. Kish,
AIP {\bf 511}, 422.
See also the related preprint cond-mat/9905086.

\bibitem{gd3}
M. P. Das and F. Green, in {\it Proceedings of the 23rd International
Workshop on Condensed Matter Theories},
ed. G. S. Anagnostatos (Nova Science, New York, 2000), in press
(preprint cond-mat/9910183).

\bibitem{gd4}
F. Green and M. P. Das, J. Phys.: Cond. Mat. {\bf 12}, 5233 (2000).

\bibitem{gd5}
F. Green and M. P. Das, J. Phys.: Cond. Mat. {\bf 12}, 5251 (2000).

\bibitem{gd6}
M. P. Das and F. Green, Aust. J. Phys., to appear
(preprint cond-mat/0005124).

\bibitem{naveh}
Y. Naveh, preprint cond-mat/9806348.

\bibitem{sw}
C. J. Stanton and J. W. Wilkins, Phys. Rev. B {\bf 35}, 9722 (1987);
{\it ibid.} {\bf 36}, 1686 (1987).
\bibitem{ldr93}
R. Landauer, Phys. Rev. B {\bf 47}, 16427 (1993).

\bibitem{dang}
D. T. Gillespie, J. Phys.: Cond. Mat. {\bf 12}, 4195 (2000).

\bibitem{pinoz}
D. Pines and P. Nozi\`eres, {\it The Theory of Quantum Liquids}
(Benjamin, New York, 1966).

\bibitem{fenton}
E. W. Fenton, Superlattices and Microstr. {\bf 16}, 87 (1994).

\end{references}
\end{document}